\documentclass{PoS}
\PoS{PoS(LAT2005)131}
\usepackage{graphicx}
\newcommand{\srm}[1]{{\textrm{\scriptsize{#1}}}}
\newcommand{\DCI}{D_\srm{CI}}

\title{First results from dynamical chirally improved fermions\footnote{for the Bern-Graz-Regensburg (BGR) collaboration} }

\ShortTitle{First results from dynamical chirally improved fermions }

\author{C. B.Lang \\
        Karl-Franzens-Universit\"at Graz, Austria\\
        E-mail: \email{christian.lang@uni-graz.at}}
\author{\speaker{Pushan Majumdar}\thanks{Supported by Fonds zur F\"orderung der Wissenschaftlichen Forschung 
in \"Osterreich, project M870-N08 (Lise-Meitner Fellowship)}\\
        Karl-Franzens-Universit\"at  Graz, Austria\\
        E-mail: \email{pushan.majumdar@uni-graz.at}}
\author{Wolfgang Ortner\thanks{Supported by Fonds zur F\"orderung der Wissenschaftlichen Forschung in \"Osterreich,
project P16310-N08}\\
        Karl-Franzens-Universit\"at Graz, Austria\\
        E-mail: \email{wolfgang.ortner@uni-graz.at}}

\abstract{
We simulate Quantum Chromodynamics in four Euclidean  dimensions with two
(degenerate mass)  flavors of dynamical  quarks. The Dirac operator is the
so-called chirally improved operator that has been studied so far in quenched 
calculations. We now present results of an implementation with the Hybrid Monte
Carlo (HMC) algorithm including   stout smearing. Our results are from an $8^3\times 
16$ lattice with tadpole improved L\"uscher-Weisz gauge action.  We present our
estimate of the lattice spacing, the pi and rho meson masses and evidence for
tunneling between different topological sectors.}

\FullConference{XXIIIrd International Symposium on Lattice Field Theory\\
                 25-30 July 2005\\
                 Trinity College, Dublin, Ireland}

\begin{document}

\section{Introduction}
The chirally improved Dirac operator is a generalized Dirac operator which
approximately obeys the Ginsparg-Wilson (GW) relation. Its construction and
implementation for QCD has been discussed elsewhere \cite{Ga01,GaHiLa00}. So far
this operator was used for light hadron spectroscopy in  quenched calculations by
the Bern-Graz-Regensburg (BGR) collaboration \cite{GaGoHa03a}. In those studies it
was found that smearing  the gauge links was important (using HYP smearing) as it
resulted in better chiral properties for the  operator, viz. the spectrum showed
less deviation from the Ginsparg-Wilson circle compared to the unsmeared case.
Therefore we decided to use one level of smearing  in our studies, too.

Usual HYP smearing is not well suited for use in HMC and
therefore we implement the recently introduced stout smearing. It is tailor made for
HMC and is differentiable. The main difference between the stout  smearing
and other kinds of smearing is the way in which the general $3\times 3$ complex
matrix  is projected back to SU(3). Instead of the usual Gram-Schmidt
orthogonalization  one takes the traceless anti-hermitian part of the matrix
(which is an element of the su(3) algebra) and then raises it to the group by the
exponential map \cite{MoPe04}. 

We now simulate full QCD (with two light fermions) by implementing the HMC-algorithm for $\DCI$ and
present here our first results. Details of the implementation and performance of
the HMC-updating are discussed in another contribution to these proceedings 
\cite{LaMaOr05a}.

\section{Setting the scale}

Our  Dirac operator follows \cite{GaHiLa00} but uses stout smearing for the gauge
configurations as part of the definition. We therefore had to reconstruct the 
the chirally improved Dirac operator appropriate 
for the combinations of gauge couplings and fermion masses used. The fermionic
force in the HMC trajectory was calculated using  the smeared chirally
improved Dirac operator \cite{LaMaOr05a}. 

\begin{figure}[tb]
\begin{center}
\includegraphics*[width=7cm,clip]{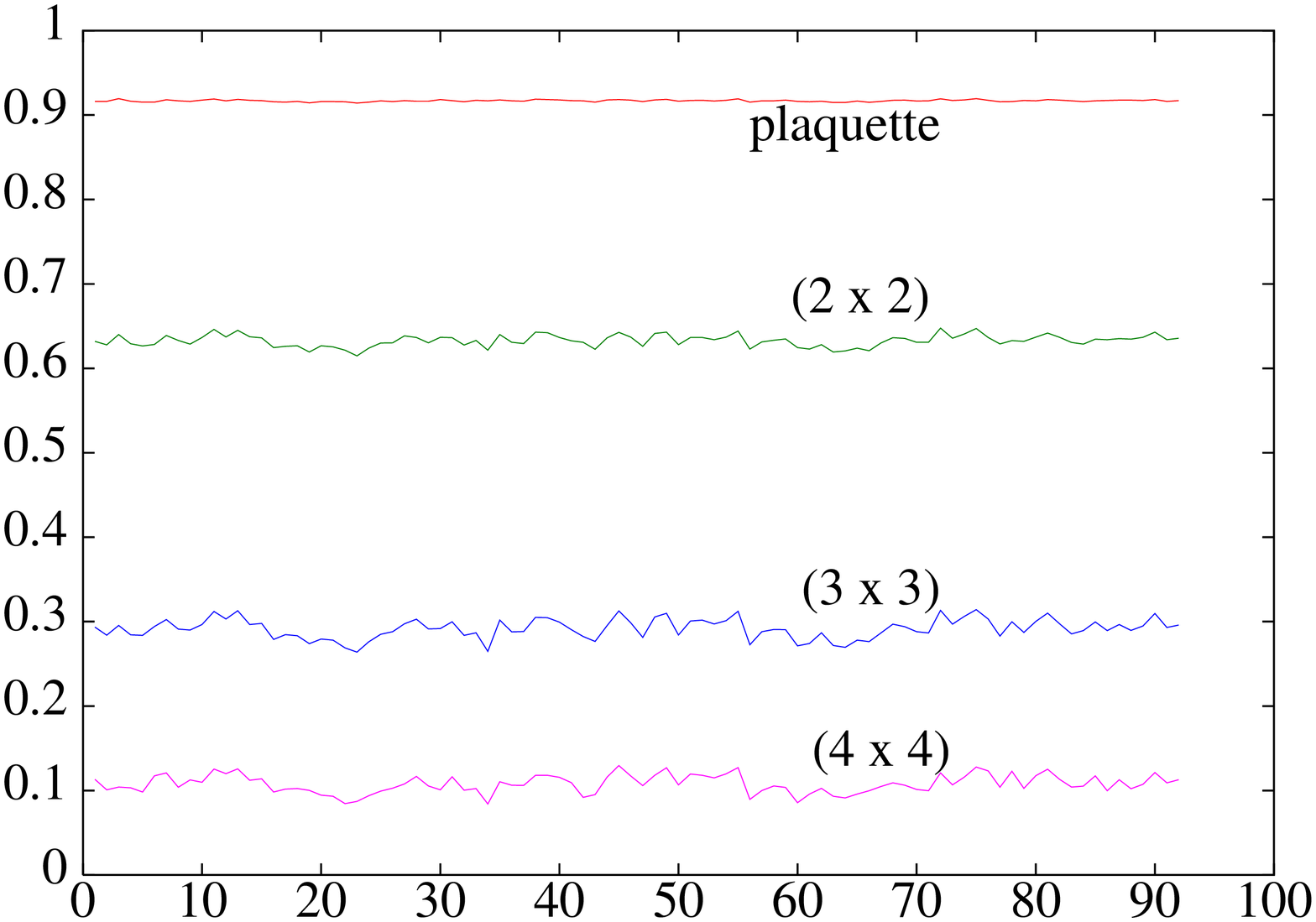}
\includegraphics*[width=7cm,clip]{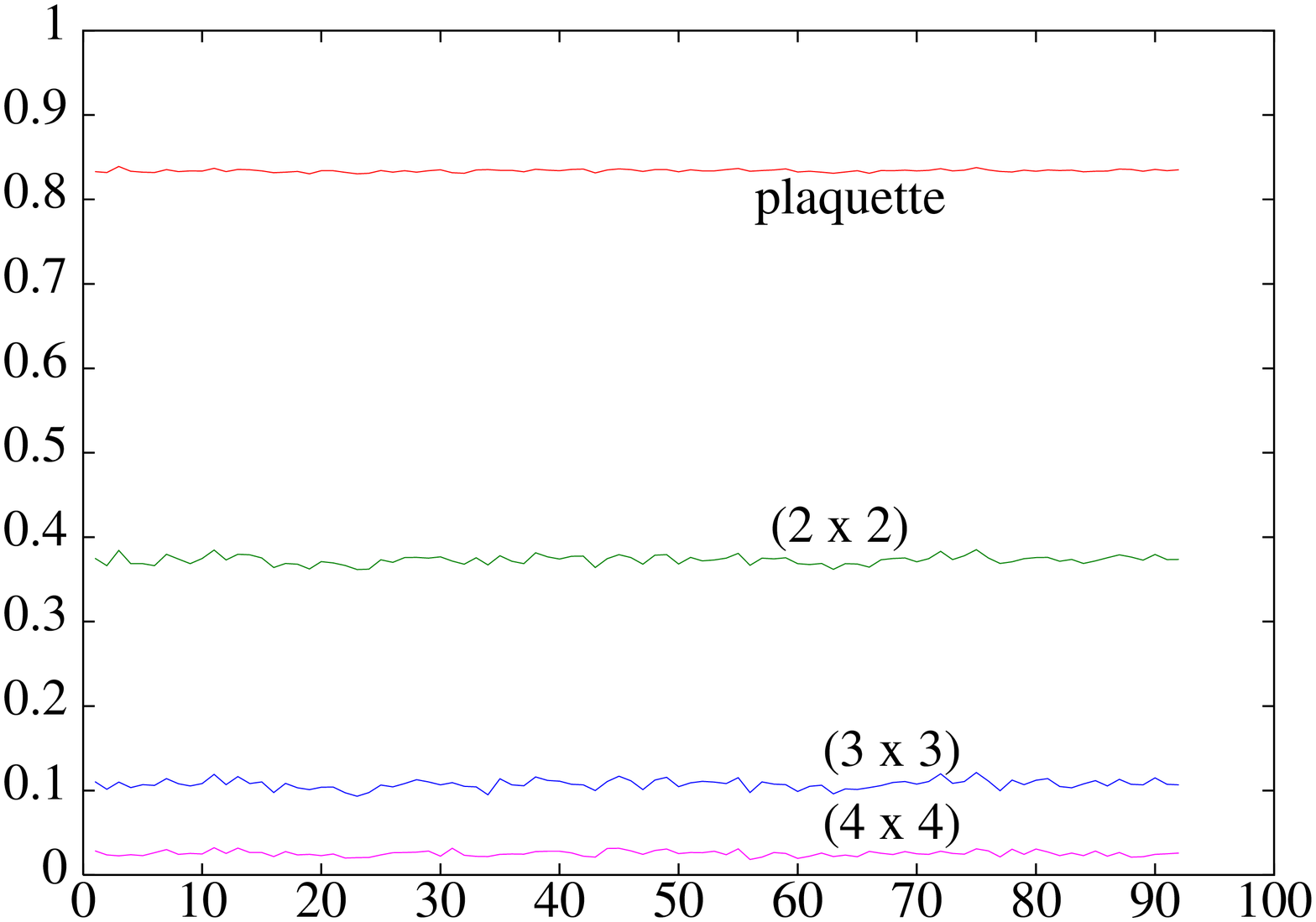}
\end{center}
\caption{HYP smeared (l.h.s.) and stout smeared (r.h.s.) Wilson loops of
different sizes. Note that for a $(4 \times 4)$ Wilson loop the expectation
value of the HYP smeared loop is $\sim$ 5  times larger than the stout smeared
one. (Data from run on line 2 in Table 1.)}
\label{loopfig}
\end{figure}

Since the stout-smearing is  (to our knowledge) being used for the first time to 
construct the $\DCI$, we find it instructive to look at the resulting difference
between HYP and stout smearing. In Fig.~\ref{loopfig} we plot different sized
Wilson loops computed from HYP-smeared and stout- smeared links obtained from
identical bare link gauge configurations.  Actual values of the plaquette are
quite sensitive to the smearing parameter for stout-smearing.  We chose this
parameter to be isotropic and one that maximizes the plaquette expectation value.
Nevertheless we found that stout-smearing has less effect on the loop
expectation values than HYP-smearing. Since smearing is a local process,
the lattice spacing, derived from long distance behavior of correlators like
loops, should not be affected by smearing. However, we expect less fluctuations 
from HYP smeared configurations.

As a first indicator we set the scale on the lattice by using the Sommer
parameter.  (We may change to using hadron masses at a later stage.) We derive
this from HYP smeared Wilson loops of various sizes. Computing the static
potential by 
\begin{equation}\label{sommer}
V(r)=-\frac{1}{T_2-T_1}\,\log \left[\frac{W(r,T_2)}{W(r,T_1)}\right] \;.
\end{equation}
and assuming the usual form of the potential $V(r) = \sigma r +\mu + c/r$, we 
obtain the parameters $\sigma,\,\mu$ and $c$ by fitting. Then the Sommer
parameter  is as usual given by $r_0^2F(r_0)=1.65$, which in this case translates
to  $r_0=\sqrt{(1.65+c)/\sigma}$. To  set the scale explicitly in fermi we use
$r_0=0.5$ fm. The sizes of the Wilson loops are   such that $r_0$ lies inside the
range of $r$ used in the fits.

\begin{figure}[t]
\begin{center}
\includegraphics*[width=8truecm]{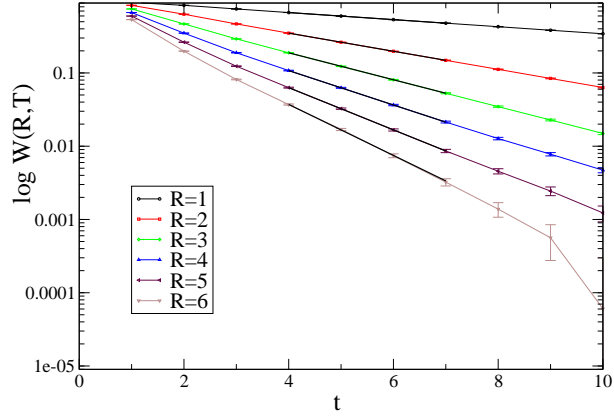}
\end{center}
\caption{ Sommer parameter ($r_0$) using HYP smeared Wilson loops.
$8^3\times 16$ lattice with lattice spacing $a=0.135$ fm 
and bare quark mass $m_q=0.05$ in units of $a$.}
\end{figure}

To make sure that the lattice spacing obtained in this way  (cf. Table~\ref{tab1})
does not depend on the smearing procedure, we repeated the calculation with stout
smearing. That reproduced the results, albeit with slightly larger error bars.

\section{Results}

\begin{table}[t]
\begin{center}
  \begin{tabular}{rlllllllllll}
    \hline
    \hline
    \(L^3 \times T\)  & \(a\,m\) &$\beta_\srm{LW}$ & $res$     &$\Delta t$&steps& acc.       &HMC   & \(a\)[fm]  \vspace{-3pt}\\
                      &          &                 &		 &	  &	&	     &time	   &		\\
    \hline
    \(8^3 \times 16\) &  0.05    & 5.4 & \(10^{-10}\) &0.015  & 50  &$\sim$93\%  &700      &   0.114(3)\\
    \(8^3 \times 16\) &  0.05    & 5.3 & \(10^{-10}\) &0.015  & 50  &$\sim$91\%  &700      &   0.135(3)\\
    \(8^3 \times 16\) &  0.08    & 5.4 & \(10^{-10}\) &0.015  & 50  &$\sim$93\%  &700      &   0.138(3)\\
    \(12^3 \times 24\)&  0.05    & 5.3 & \(10^{-8}\)   &0.01  & 100 &$\sim$87\%  &200      &   0.129(3) \\
     \hline
  \end{tabular}
  \caption{\label{tab1}Parameters, statistics and some results from our dynamical simulations.}
\end{center}
\end{table}

In Table~\ref{tab1} we summarize our run parameters. More details are given in
\cite{LaMaOr05a}.

Next we look at the correlators of the two lightest mesons,  the $\pi$ and the
$\rho$ mesons. Here we measure only point-to-point correlators defined by
\begin{eqnarray}
C_{\pi}(0,t) &=& \sum_{\vec x}\; \textrm{tr} \left( \gamma_5 \,\DCI^{-1}({\vec x},\,t:0,\,0)\, \gamma_5 \,\DCI^{-1}(0,\,0:{\vec 
x},\,t)\right)\;, \\
C_{\rho}(0,t) &=& \sum_{{\vec x}, i=1,2,3} \,\;\textrm{tr} \left( \gamma_i \;\DCI^{-1}({\vec 
x},t:0,\,0)\, \gamma_i \,\DCI^{-1}(0,\,0:{\vec x},\,t)\right)\;.
\end{eqnarray}

\begin{figure}[t]
\begin{center}
\includegraphics[width=7.6cm,angle=-90]{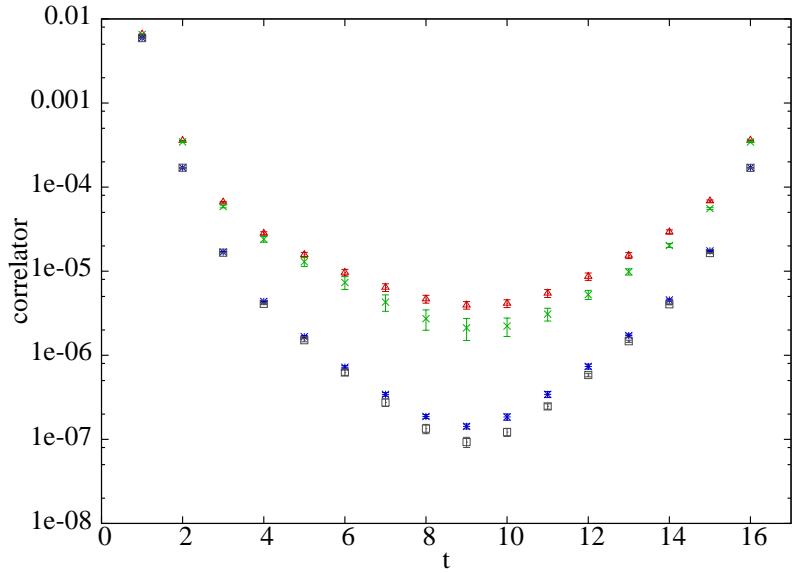}
\end{center}
\caption{$\rho\,(\ast\,,\,\Box)$ and $\pi\,(\triangle\,,\,\times)$ correlators on
$8^3\times 16$  lattices ($a\,m_q=0.05$). The upper line corresponds to lattice
spacing $a=0.135$ fm while the lower one is for $a=0.114$ fm in both channels.} 
\label{propfig}
\end{figure}

To extract the masses, the data were folded about the symmetry point and fitted to
the functional  form $a\;\cosh[m(t-T/2)]$. For the $8^3\times 16$ lattices, there
seemed to be some contamination from the higher states and it was not clear if we
were indeed observing the asymptotic decays for the correlators  (cf.
Fig.~\ref{propfig}). On  $12^3\times 24$ lattices, such contaminations were less
pronounced and within error bars we seem to obtain the asymptotic decays in the
t-ranges shown in the Table \ref{masstab}.  In this table we report on the masses
for the $\pi$ and the $\rho$ meson as obtained from these correlators determined
from fits in the given regions. The errors were obtained  from a jackknife
analysis. Due to the simple point-like sources the small distance behavior is
dominated by excited states. This can be  improved by using smeared sources.

Since the $\DCI$ is only approximately a Ginsparg-Wilson  operator, we have to  a
posteriori confirm that the good chiral properties that we expect are indeed 
present in the operator. As a cross-check for that we compare the eigenvalue
spectrum of $\DCI$ for several configurations with dynamical fermions with those of
a quenched simulation. The spectra have similar ``fuzzyness''  and are all close
to the unit circle, as shown in \cite{LaMaOr05a}.

For Dirac operators obeying the $\gamma_5$ hermiticity, real eigenmodes are the
carriers of non-vanishing  chirality  and enumerate the total net topological
charge. For massless exact GW fermions the lowest-lying eigenvalues are all
exactly at the zero of the complex plane. For our approximate GW operator, even in
the massless case, the lowest-lying eigenvalues are not exactly at zero but have a
small real offset.  Their total number defines the net topological charge.

\begin{table}[t]
\begin{center}
\begin{tabular}{ccccclcl}
\hline
\hline
lattice-size&$a$(fm)&bare $m_q$ &{t - range} & {$a\,m_{\pi}$} & $\chi^2/d.o.f$ & {$a\,m_{\rho}$} & $\chi^2/d.o.f$ \\
\hline
$8^3\times 16$ & 0.135(3) & 0.05 & {5-8} & {0.520(15)} & $\sim$ 0.4 & {0.808(14)} & $\sim$ 1.5 \\
               & &&{6-8} & {0.500(15)} &  $\sim$ 0.05 & {0.776(15)} & $\sim$ 0.1 \\
&&&&&&&\\
$8^3\times 16$ & 0.114(3) & 0.05 &{5-8}& {0.598(58)} & $\sim$ 0.06 & {0.881(26)} & $\sim$ 0.7 \\
               & & &{6-8}& {0.589(64)} & $\sim$ 0.02 & {0.858(28)} & $\sim$ 0.06 \\
&&&&&&&\\
$8^3\times 16$ & 0.138(3) & 0.08 &{5-8}& {0.621(18)} & $\sim$ 0.1 & {0.860(13)} & $\sim$ 1.3 \\
               & &&{6-8} & {0.611(23)} & $\sim$ 0.01 & {0.839(17)} & $\sim$ 0.1 \\
&&&&&&&\\
$12^3\times 24$ & 0.129(3) & 0.05 &{7-10}& {0.384(21)} & $\sim$ 0.08 & {0.575(37)} & $\sim$ 0.01 \\
                & & &{8-10}& {0.366(26)} & $\sim$ 0.04 & {0.575(15)} & $\sim$ 0.06 \\
\hline
\end{tabular}
\end{center}
\caption{$\pi$ and $\rho$ masses (in lattice units) from point to point correlators}
\label{masstab}
\end{table}

\begin{figure}[t]
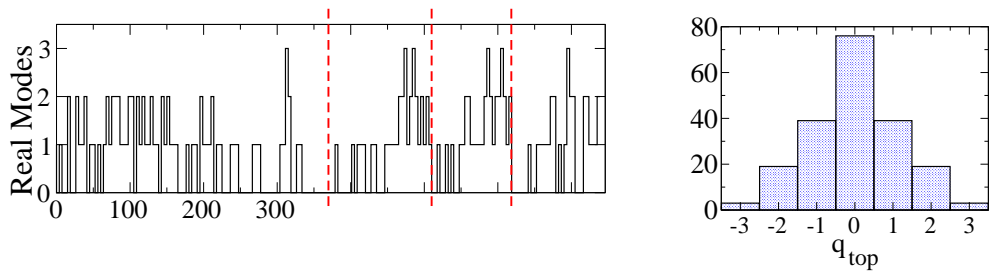

\begin{center}
  \includegraphics*[width=8cm]{tophist_5.3.eps}\hspace{10mm}
  \includegraphics*[width=4cm]{top_histogram.eps}
\end{center}
\caption{History of real modes and symmetrized histogram of the topological charge.}
\label{topfig}\end{figure}

It is important to determine the amount of autocorrelation present in the system.
This is explored in \cite{LaMaOr05a}. Here we look at another  
closely related question, whether the algorithm
samples all topological sectors of our model in the right way. We show a history of the
number of close-to-zero modes (i.e., the absolute value of the topological charge) and 
symmetrized histogram of the topological charge (it is symmetrized simply because we did
not compute the chirality of our real modes) in Fig. ~\ref{topfig}. Data taken from different Markov
chains are separated by dashed lines in the figure. We do observe tunneling (which
seems to be a problem for the overlap operator) and a Gaussian-like shape of the
distribution.

\begin{figure}[t]
\begin{center}
  \includegraphics*[width=6cm]{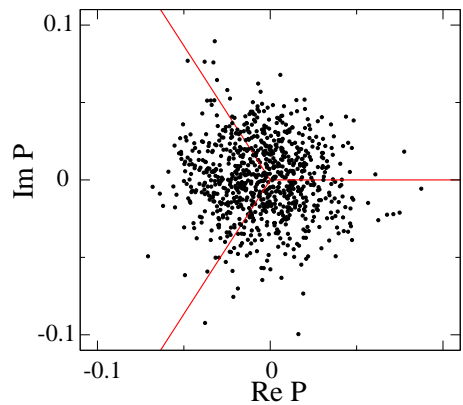}
\end{center}
\caption{Distribution of ``spatial" Wilson lines for $8^3\times 16$ lattice at $a=$0.135 fm}
\label{polfig}
\end{figure}
\section{Conclusions}

We have successfully implemented HMC with the chirally improved Dirac operator and
our  first results seem to indicate that the good chiral properties which were
observed in the  quenched case seem to be present in the dynamical case also. In
contrast to the overlap operator, we do not have a strong barrier at the
topological boundaries and we see quite often evidence of tunneling, manifested as
``zero modes'' of the $\DCI$. We also confirm (Fig.~\ref{polfig}) that we have not
undergone spatial deconfinement by looking at Wilson  lines along any one of the
shorter directions of the lattice. 

We estimate the masses of the $\pi$ and $\rho$ meson using point-to-point
correlators. From quenched calculations \cite{GaGoHa03a} we know that $8^3\times
16$ lattices (at $a=$0.15 fm) still have substantial size dependence and we expect
the same to hold here too. However, the $12^3\times 24$ lattices ($\sim
1.5$ fm) should be better behaved. In this exploratory study we are at
comparatively heavy quark masses $m_{\pi}/m_{\rho}\approx 0.6$. Runs using lighter
masses are in progress. 



\begin{thebibliography}{99}

\bibitem{Ga01}
C.~Gattringer, \emph{A new approach to Ginsparg-Wilson fermions}, Phys. Rev. D {\bf 63}
  (2001) 114501 [{\tt hep-lat/0003005}].

\bibitem{GaHiLa00}
C.~Gattringer, I.~Hip, and C.~B. Lang, \emph{Approximate Ginsparg-Wilson fermions: A
  first test}, Nucl. Phys. B {\bf 597} (2001) 451 [{\tt hep-lat/0007042}].

\bibitem{GaGoHa03a}
C.~Gattringer, M.~G{\"o}ckeler, P.~Hasenfratz, et.~al., 
\emph{Quenched spectroscopy with fixed-point and chirally improved
  fermions}, Nucl. Phys. B {\bf 677} (2004) 3 [{\tt hep-lat/0307013}].

\bibitem{MoPe04}
C.~Morningstar and M.~Peardon, \emph{Analytic smearing of SU(3) link variables in
  lattice QCD}, Phys.Rev. D {\bf 69} (2004) 054501 [{\tt hep-lat/0311018}].

\bibitem{LaMaOr05a}
C.~B. Lang, P.~Majumdar and W.~Ortner, 
\emph{Implementing dynamical chirally improved
  fermions in lattice qcd}, these proceedings, PoS(LAT2005)124.

\end{thebibliography}
\end{document}